# Sunspots during the Maunder Minimum from *Machina Coelestis* by Hevelius

V.M.S. Carrasco[1,2], J. Villalba Álvarez[3] and J.M. Vaquero[4]

[1] Centro de Geofísica de Évora, Universidade de Évora, Évora, Portugal.

[2] Departamento de Física, Universidad de Extremadura, Badajoz, Spain.

[3] Departamento de Ciencias de la Antigüedad, Universidad de Extremadura, Cáceres, Spain.

[4] Departamento de Física, Universidad de Extremadura, Mérida, Spain [e-mail: jvaquero@unex.es].

**Abstract:** We revisited the sunspot observations published by Johannes Hevelius in his book *Machina Coelestis* (1679) corresponding to the period 1653-1675 (just in the middle of the Maunder Minimum). We show detailed translations of the original Latin texts describing the sunspot records and provide the general context of these sunspot observations. From this source only, we present an estimate of the annual values of the Group Sunspot Number based only on the records that explicitly inform about the presence or absence of sunspots. Although we obtain very low values of the Group Sunspot Number, in accordance with a grand minimum of solar activity, these values are significantly higher in general than the values provided by Hoyt and Schatten (1998) for the same period.

## 1. Introduction

The Maunder Minimum (MM) is the period spanning the years 1645 to 1715 whose main feature was prolonged low solar activity (Eddy, 1976; Soon and Yaskell, 2003). It has been the only Grand Minimum that has occurred during the telescopic era, and therefore, the study of this event has a great importance in Solar Physics and Geophysics for a better understanding of the behavior of long-term solar activity and its influence on the Earth's climate. Thus, it is important to recover the largest number of high-quality solar observations during this period in order to improve our knowledge of this phenomenon (Vaquero and Vázquez, 2009).

The recovery of thousands of solar observations made during the MM, especially by Hoyt and Schatten (1998), allows one to obtain a general scenario characterized by the following facts:

(i) The MM was well covered by direct sunspot observations according to Hoyt and Schatten (1996). One has solar observations for more than 95% of days.

(ii) Sunspots appeared rarely (during ∼ 2% of the days), without any indication of the 11-year cycle, which became visible only in the late phase of the Maunder minimum (Ribes and Nesme-Ribes, 1993).

(iii) Transition from the normal high activity to the deep minimum was sudden (within a few years) and without any apparent precursor. However, the recovery to the normal activity level was gradual, taking several decades.

(iv) There was a strong north-south asymmetry during the MM. In particular sunspots were only observed in the southern solar hemisphere at the end of the Maunder minimum (Ribes and Nesme-Ribes, 1993).

In recent years, several studies have been changing some of our ideas of these characteristics of the MM. Vaquero et al. (2011) carried out a review of the Group sunspot number (GSN) in the years previous to the onset of MM, showing reduced solar activity in two previous cycles before the start of the Minimum. This would mean a possible gradual transition between prolonged high and low solar activity. While it is almost impossible to detect the 11-year periodicity of the solar cycle in MM, the results obtained by Poliuanov et al. (2014) show that cosmic rays may have exhibited periods of 11 years in this period. Clette et al. (2014) have shown that a large number of sunspot observations during the MM recovered by Hoyt and Schatten (1996, 1998) were obtained from solar meridian observations. This fact could imply that solar activity during the MM is underestimated. In particular, Vaquero and Gallego (2014), after an evaluation of the preserved records of solar meridian observations (accompanied by notes regarding sunspots) made at the Royal Observatory of the Spanish Navy (San Fernando, Spain) during the period 1833-1840, concluded that such records should be used with extreme caution for the reconstruction of past solar activity. Recently, Vaquero and Trigo (2015) proposed a redefinition of the extent of MM (commonly established from 1645 to 1715) by dividing it into three periods: a central region called the Deep Maunder Minimum which spans from 1645 until 1700, and two transition regions, 1618-1645 and 1701-1723.

In this paper, we hope to contribute to knowledge of the solar activity during the MM by presenting a modern translation of the sunspot observations published by Hevelius in his book *Machina Coelestis Pars Posterior* (Hevelius, 1679), including a quantitative analysis of these records. In Section 2, we describe the solar observations recorded in this source. Estimation of the values of the GSN from these observations is performed in Section 3. Finally, the conclusions are drawn in Section 4. We include an Appendix containing the original Latin texts and their corresponding translations into modern English.

## 2. Sunspot Data in *Machina Coelestis*

Johannes Hevelius published in the last part of his work *Machina Coelestis* a large number of solar observations. This record was studied previously by Hoyt and Schatten (1995) and Rek (2013) from the point of view of the reconstruction of solar activity. The two studies diverge in their analysis and results. Table 1 lists the annual number of active days (AD, the number of days with spots on the solar disc), quiet days (QD, the number of days without spots on the solar disc), and total days with a record (N) recorded in *Machina Coelestis* according to Hoyt and Schatten (1995) and Rek (2013). The differences between the two interpretations are remarkable. The numbers of total and quiet days provided by Hoyt and Schaten (1995) are much higher than those provided by Rek (2013). However, the annual values of active days provided by the two publications are nearly equal. Our aim is, after a careful study including the translation of the original Latin texts, provide a definitive result for this source understanding the origin of these differences.

Hevelius listed his measurements of solar meridian altitude in a table with six columns. Figure 1 shows, as an example, the Page 8 of the last part of the book *Machina Coelestis*. The first column shows the date of the record. The meridian altitude of the Sun is listed in the second column. The following columns indicate different information about the instrument used in the observation (third column), the meteorological conditions (fourth column), the quality (fifth column), and notes (last column). It is just in this last column where Hevelius made his annotations about sunspot observations.

Hoyt and Schatten (1998) used all the observation dates of solar meridian altitude to fill their sunspot database. If they did not find any indication about the presence or absence of sunspots in the comments column, then they assumed that there were no sunspots on the solar disc. Therefore, there are very more days of observation in the Hoyt and Schatten database than in subsequent works (Rek, 2013 and this work). Indeed, we should note that the explicit sunspot data recorded in *Machina Coelestis* are not associated with the solar meridian observations. As an example, one detects four different cases in Figure 2 that support this statement: the red arrow points to a case (23 February 1660) in which there is one sunspot record and there is also one measurement of the solar meridian altitude; the green arrow indicates a day (24 February 1660) on which there is one sunspot record but there is no solar meridian observation; the blue arrow points to a day (1 March 1660) with one observation of no spots on the solar disc (quiet day) and without a solar meridian observation; finally, the orange arrow shows a day (12 March 1660) with one record of no spots on the solar disc (quiet day) and with one measurement of the solar meridian altitude. Therefore, the two types of observation (solar meridian altitude and solar disc) are not closely associated and probably the instruments used for the two observations were quite different.

In this paper, we check carefully the table of solar meridian altitudes published by Hevelius in the last part of his book *Machina Coelestis*, extracting the explicit information about the presence or absence of sunspot. The original texts were transcribed and translated from Latin to English. We provide the original and translated texts in the Appendix of this paper. We must emphasize that we have rewritten the text in standard Latin, i.e., correcting accents, expanding abbreviations, etc. The table of this Appendix is divided into four columns. The first column lists the year of the observation. The second column lists the day and the month. The third column contains the transcription in Latin of the original text with its corresponding translation in brackets. Finally, we have added a column indicating the character of this date: (AD) active or (QD) quiet day.

## 3. Estimation of sunspot number with the observations of Hevelius

Several studies (Kovaltsov et al., 2004; Vaquero et al., 2012) have been carried out in order to assess the past solar activity from the estimated number of active days. In order to estimate annual values of the GSN index from the sunspot records presented in the Appendix, we have followed the expression proposed by Kovaltsov et al. (2004) to estimate the GSN index from the fraction of active days ($F_a = N_a/N$), where $N_a$ is the number of days with sunspot observations and N is the number of days of a year. This expression is the following power law: $GSN = 19 \cdot F_a^{1.25}$. This relation between GSN and $F_a$ works well for low solar activity but not for medium and high solar activity (GSN > 30). Table 2 shows the annual number of active, quiet, and total days recorded in *Machina Coelestis* and the expected values for the fraction of active days (according to the texts included in the Appendix) for the different years in which Hevelius recorded explicitly sunspot observations. Moreover, values of the fraction of active days computed by Hoyt and Schatten (1998) ($F_{a\text{-}HS98}$) and Rek (2013) ($F_{a\text{-}R13}$) are listed for comparison. Our estimate of the fraction of active days agrees with the results of Rek (2013) in general but our values are always equal to or less than the values given by Rek (2013). The values provided by Hoyt and Schatten (1998) are very small due to the high number of quiet days obtained from the measurement dates of solar meridian altitudes, when explicit information on sunspots is not available. It is worth noting that Hoyt and Schatten (1998) only used this kind of "astrometric observations" for early time-periods when few sunspot observations are available. Therefore, this problem does not affect their solar activity reconstruction in the rest of the time-periods (18th, 19th, and 20th centuries).

Figure 3 shows the fraction of active days (%) (upper panel) and the GSN (lower panel) during the period 1653-1675. We use red symbols for the values obtained in this work and blue symbols for the data published by Hoyt and Schatten (1998). We include error bars represent a 99% confidence interval. We computed these error bars by assuming a hypergeometric probability distribution for these values. Note that we are implicitly assuming that the solar disc observations by Hevelius are random and independent of each other. We accept that this is not exactly true because the observations might be more numerous during the days with sunspots. However, we can thus offer a first slightly overestimated $F_a$ and GSN during the MM. If one wants to obtain the uncertainty of the number of active days in a year from a sample of *n* observations with *r* active days, one must use the well-known expression of the hypergeometric probability distribution:

$$p(s) = \frac{s!(N-s)!}{(s-r)!(N-s-n+r)!} \cdot \frac{n!(N-n)!}{(n-r)!N!r!}$$

where *N* is the number of days in a year (365 or 366) and *s* is the total number of active days within the year which is to be estimated. Using this distribution, we computed the most probable value of *s* together with the 99% confidence interval (Figure 3). The red and blue solid lines in the lower panel represent the mean values of GSN obtained in this work and by Hoyt and Schatten (1998), respectively. In order to compute these mean values, we used the power law given by Kovaltsov et al. (2004) and the expected value of active days (%) during the complete period (1653-1675). Note that the GSN values provided by Hoyt and Schatten (1998) corresponding to the years 1653, 1654, 1657, and 1660 are outside our 99% confidence interval (Figure 3, lower panel).

Rek (2013) used the solar observations by Hevelius to claim that the Maunder Minimum is unreal. We deeply disagree with this statement. Our estimates of the fraction of active days during the period 1653-1675 using the solar observations by Hevelius are very low (see Table 3) in comparison with the values of this parameter during the Dalton Minimum (not considered to be a Grand Minimum, but a period of reduced solar activity). Thus, our values of GSN estimated from the observations made by Hevelius (much higher than those given by Hoyt and Schatten, 1998) clearly indicate a period of Grand Minimum.

## 4. Conclusions

We have carefully analysed the explicit information about sunspot recorded by Johannes Hevelius in his book *Machina Coelestis* during the period 1653-1675, just in the core of the Maunder Minimum. We include an Appendix with the original transcriptions and the modern translations of the explicit information on sunspot available in this historical source. From these texts we have evaluated the observation days as active or quiet. We have obtained the annual values of the fraction of active days, and have estimated the corresponding GSN for the whole period. We have obtained significantly greater values than the values provided in the seminal work of Hoyt and Schatten (1998), who used days without explicit sunspot observation but with measurements of solar meridian altitude to fill their database with zero values. We think that some GSN values given by Hoyt and Schatten (1998) might be underestimates. Therefore, a deep revision of the GSN during the MM needs to be undertaken.

**Acknowledgements**

Support from the Junta de Extremadura (Research Group Grant No. GR10131), from the Ministerio de Economía y Competitividad of the Spanish Government (AYA2011-25945), and from the COST Action ES1005 TOSCA (http://www.tosca-cost.eu) is gratefully acknowledged.

**Table 1.** Annual number of active days (AD), quiet days (QD), and total days with a record (N) recorded in *Machina Coelestis* according to Hoyt and Schatten (1995) and Rek (2013).

| | Hoyt and Schatten (1998) | | | Rek (2013) | | |
|---|---|---|---|---|---|---|
| Year | AD | QD | N | AD | QD | N |
| 1653 | 11 | 81 | 92 | 12 | 10 | 22 |
| 1654 | 4 | 67 | 71 | 4 | 0 | 4 |
| 1657 | 4 | 35 | 39 | 4 | 2 | 6 |
| 1658 | 0 | 23 | 23 | - | - | - |
| 1659 | 0 | 56 | 56 | 0 | 46 | 46 |
| 1660 | 30 | 66 | 96 | 28 | 30 | 58 |
| 1661 | 3 | 70 | 73 | 2 | 17 | 19 |
| 1663 | 0 | 65 | 65 | 0 | 7 | 7 |
| 1671 | 5 | 67 | 72 | 2 | 3 | 5 |
| 1675 | 0 | 8 | 8 | 0 | 2 | 2 |

**Table 2.** Annual number of active, quiet, and total days recorded in *Machina Coelestis*, and expected value for the fraction of active days (according to the texts included in the Appendix). Additionally, values of the fraction of active days computed by Hoyt and Schatten (1998) ($F_{a\text{-}HS98}$) and Rek (2013) ($F_{a\text{-}R13}$) are listed for comparison.

| YEAR | ACTIVE | QUIET | TOTAL | $F_a$ | $F_{a\text{-}HS98}$ | $F_{a\text{-}R13}$ |
|---|---|---|---|---|---|---|
| **1653** | 11 | 11 | 22 | 0.50 | 0.07 | 0.55 |
| **1654** | 3 | 1 | 4 | 0.75 | 0.03 | 1.00 |
| **1657** | 3 | 3 | 6 | 0.50 | 0.01 | 0.67 |
| **1658** | 0 | 2 | 2 | 0.00 | 0.00 | - |
| **1659** | 0 | 47 | 47 | 0.00 | 0.00 | 0.00 |
| **1660** | 28 | 31 | 59 | 0.47 | 0.16 | 0.48 |
| **1661** | 2 | 17 | 19 | 0.11 | 0.06 | 0.11 |
| **1663** | 0 | 7 | 7 | 0.00 | 0.00 | 0.00 |
| **1671** | 2 | 3 | 5 | 0.40 | 0.08 | 0.40 |
| **1675** | 0 | 2 | 2 | 0.00 | 0.00 | 0.00 |

**Table 3.** Expected, upper limit, and lower limit (99% confidence interval) values of the fraction of active days (%) during the MM (from the observations by Hevelius analysed here from 1653 to 1675) and during the Dalton Minimum (from the seminal work of Hoyt and Schatten, 1998, from 1799 to 1821).

| Source | Expected value | Upper limit | Lower limit |
|---|---|---|---|
| HEVELIUS (1653-1675) | 28.32% | 34.55% | 22.09% |
| HS98 (1799-1821) | 50.73% | 51.51% | 49.94% |

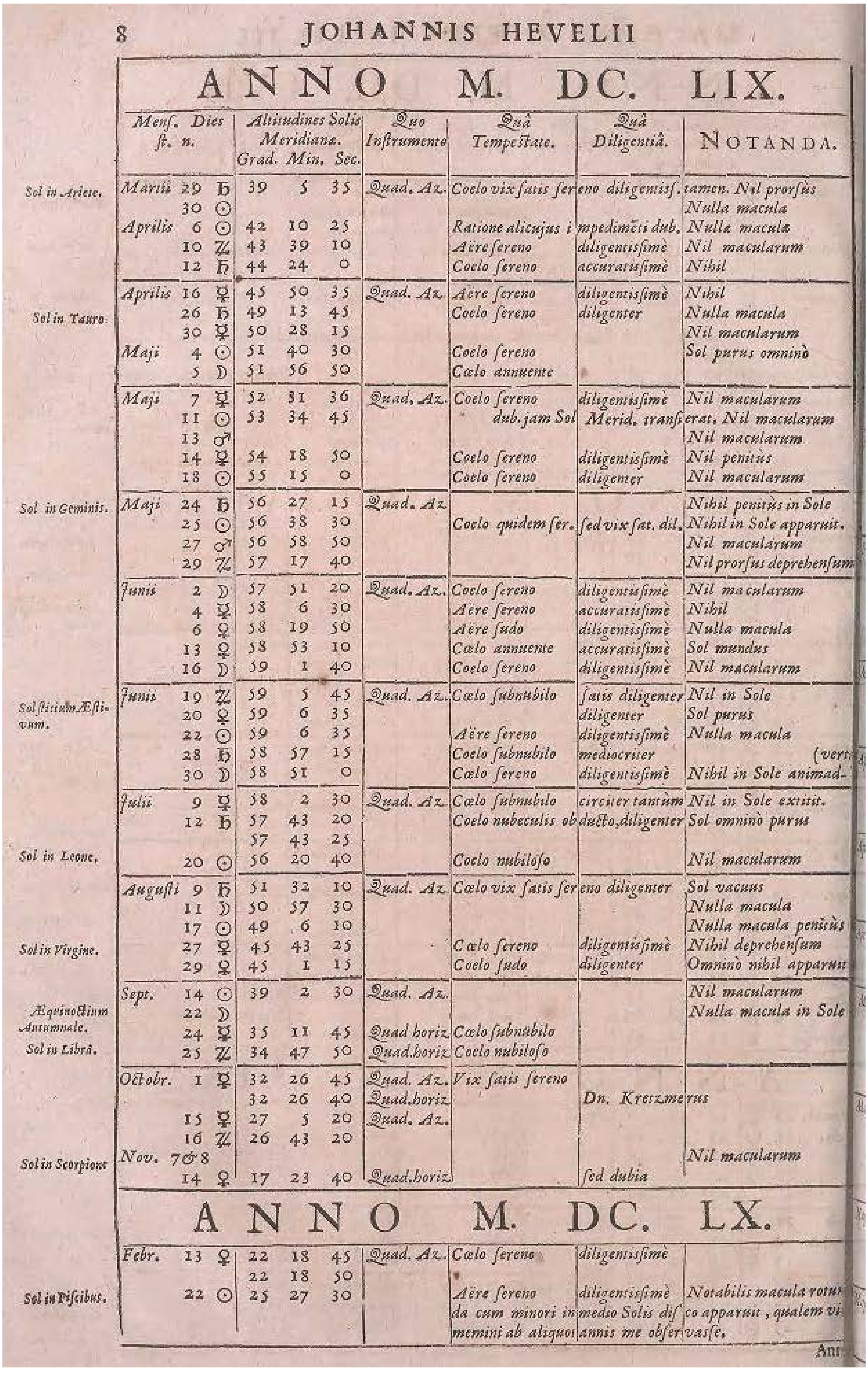

8 JOHANNIS HEVELII

ANNO M. DC. LIX.

| | Menſ. | Dies n. | ſt. | Altitudines Solis Meridianæ. Grad. | Min. | Sec. | Quo Inſtrumento | Quâ Tempeſtate. | Quâ Diligentiâ. | NOTANDA. |
|---|---|---|---|---|---|---|---|---|---|---|
| Sol in Ariete. | Martii | 29 | ♄ | 39 | 5 | 35 | Quad. Az. | Coelo vix ſatis ſer | eno diligentiſſ. | tamen. Nil prorſus |
| | | 30 | ☉ | | | | | | | Nulla macula |
| | Aprilis | 6 | ☉ | 42 | 10 | 25 | | Ratione alicujus i | mpedimẽti dub. | Nulla macula |
| | | 10 | ♃ | 43 | 39 | 10 | | Aëre ſereno | diligentiſſimè | Nil macularum |
| | | 12 | ♄ | 44 | 24 | 0 | | Coelo ſereno | accuratiſſimè | Nihil |
| | Aprilis | 16 | ☿ | 45 | 50 | 35 | Quad. Az. | Aëre ſereno | diligentiſſimè | Nihil |
| Sol in Tauro. | | 26 | ♄ | 49 | 13 | 45 | | Coelo ſereno | diligenter | Nulla macula |
| | | 30 | ☿ | 50 | 28 | 15 | | | | Nil macularum |
| | Maji | 4 | ☉ | 51 | 40 | 30 | | Coelo ſereno | | Sol purus omninò |
| | | 5 | ☽ | 51 | 56 | 50 | | Cœlo annuente | | |
| | Maji | 7 | ☿ | 52 | 31 | 36 | Quad. Az. | Coelo ſereno | diligentiſſimè | Nil macularum |
| | | 11 | ☉ | 53 | 34 | 45 | | dub. jam Sol | Merid. tranſi | erat. Nil macularum |
| | | 13 | ♂ | | | | | | | Nil macularum |
| | | 14 | ☿ | 54 | 18 | 50 | | Coelo ſereno | diligentiſſimè | Nil penitus |
| | | 18 | ☉ | 55 | 15 | 0 | | Coelo ſereno | diligenter | Nil macularum |
| Sol in Geminis. | Maji | 24 | ♄ | 56 | 27 | 15 | Quad. Az. | | | Nihil penitus in Sole |
| | | 25 | ☉ | 56 | 38 | 30 | | Coelo quidem ſer. | ſed vix ſat. dil. | Nihil in Sole apparuit. |
| | | 27 | ♂ | 56 | 58 | 50 | | | | Nil macularum |
| | | 29 | ♃ | 57 | 17 | 40 | | | | Nil prorſus deprehenſum |
| | Junii | 2 | ☽ | 57 | 51 | 20 | Quad. Az. | Coelo ſereno | diligentiſſimè | Nil macularum |
| | | 4 | ☿ | 58 | 6 | 30 | | Aëre ſereno | accuratiſſimè | Nihil |
| | | 6 | ♀ | 58 | 19 | 50 | | Aëre ſudo | diligentiſſimè | Nulla macula |
| | | 13 | ♀ | 58 | 53 | 10 | | Cœlo annuente | accuratiſſimè | Sol mundus |
| | | 16 | ☽ | 59 | 1 | 40 | | Coelo ſereno | diligentiſſimè | Nil macularum |
| Solſtitium Æſtivum. | Junii | 19 | ♃ | 59 | 5 | 45 | Quad. Az. | Cœlo ſubnubilo | ſatis diligenter | Nil in Sole |
| | | 20 | ♀ | 59 | 6 | 35 | | | diligenter | Sol purus |
| | | 22 | ☉ | 59 | 6 | 35 | | Aëre ſereno | diligentiſſimè | Nulla macula |
| | | 28 | ♄ | 58 | 57 | 15 | | Coelo ſubnubilo | mediocriter | (vert |
| | | 30 | ☽ | 58 | 51 | 0 | | Cœlo ſereno | diligentiſſimè | Nihil in Sole animad- |
| | Julii | 9 | ☿ | 58 | 2 | 30 | Quad. Az. | Cœlo ſubnubilo | circiter tantùm | Nil in Sole extitit. |
| | | 12 | ♄ | 57 | 43 | 20 | | Coelo nubeculis ob | ducto, diligenter | Sol omninò purus |
| | | | | 57 | 43 | 25 | | | | |
| Sol in Leone. | | 20 | ☉ | 56 | 20 | 40 | | Coelo nubiloſo | | Nil macularum |
| | Auguſti | 9 | ♄ | 51 | 32 | 10 | Quad. Az. | Cœlo vix ſatis ſer | eno diligenter | Sol vacuus |
| | | 11 | ☽ | 50 | 57 | 30 | | | | Nulla macula |
| | | 17 | ☉ | 49 | 6 | 10 | | | | Nulla macula penitùs |
| Sol in Virgine. | | 27 | ☿ | 45 | 43 | 25 | | Cœlo ſereno | diligentiſſimè | Nihil deprehenſum |
| | | 29 | ♀ | 45 | 1 | 15 | | Coelo ſudo | diligenter | Omninò nihil apparuit |
| | Sept. | 14 | ☉ | 39 | 2 | 30 | Quad. Az. | | | Nil macularum |
| Æquinoctium Autumnale. | | 22 | ☽ | | | | | | | Nulla macula in Sole |
| | | 24 | ☿ | 35 | 11 | 45 | Quad horiz. | Cœlo ſubnubilo | | |
| Sol in Librâ. | | 25 | ♃ | 34 | 47 | 50 | Quad. horiz. | Coelo nubiloſo | | |
| | Octobr. | 1 | ☿ | 32 | 26 | 45 | Quad. Az. | Vix ſatis ſereno | | |
| | | | | 32 | 26 | 40 | Quad. horiz. | | Dn. Kretzme | rus |
| | | 15 | ☿ | 27 | 5 | 20 | Quad. Az. | | | |
| | | 16 | ♃ | 26 | 43 | 20 | | | | |
| Sol in Scorpione | Nov. | 7 & 8 | | | | | | | | Nil macularum |
| | | 14 | ♀ | 17 | 23 | 40 | Quad. horiz. | | ſed dubia | |

ANNO M. DC. LX.

| | Menſ. | Dies n. | ſt. | Grad. | Min. | Sec. | Quo Inſtrumento | Quâ Tempeſtate. | Quâ Diligentiâ. | NOTANDA. |
|---|---|---|---|---|---|---|---|---|---|---|
| | Febr. | 13 | ♀ | 22 | 18 | 45 | Quad. Az. | Cœlo ſereno | diligentiſſimè | |
| | | | | 22 | 18 | 50 | | | | |
| Sol in Piſcibus. | | 22 | ☉ | 25 | 27 | 30 | | Aëre ſereno | diligentiſſimè | Notabilis macula rotun |
| | | | | | | | | da cum minori in | medio Solis diſ | co apparuit, qualem vi |
| | | | | | | | | memini ab aliquot | annis me obſer | vaſſe. |

Ann

**Figure 1.** An example page of *Machina Coelestis*.

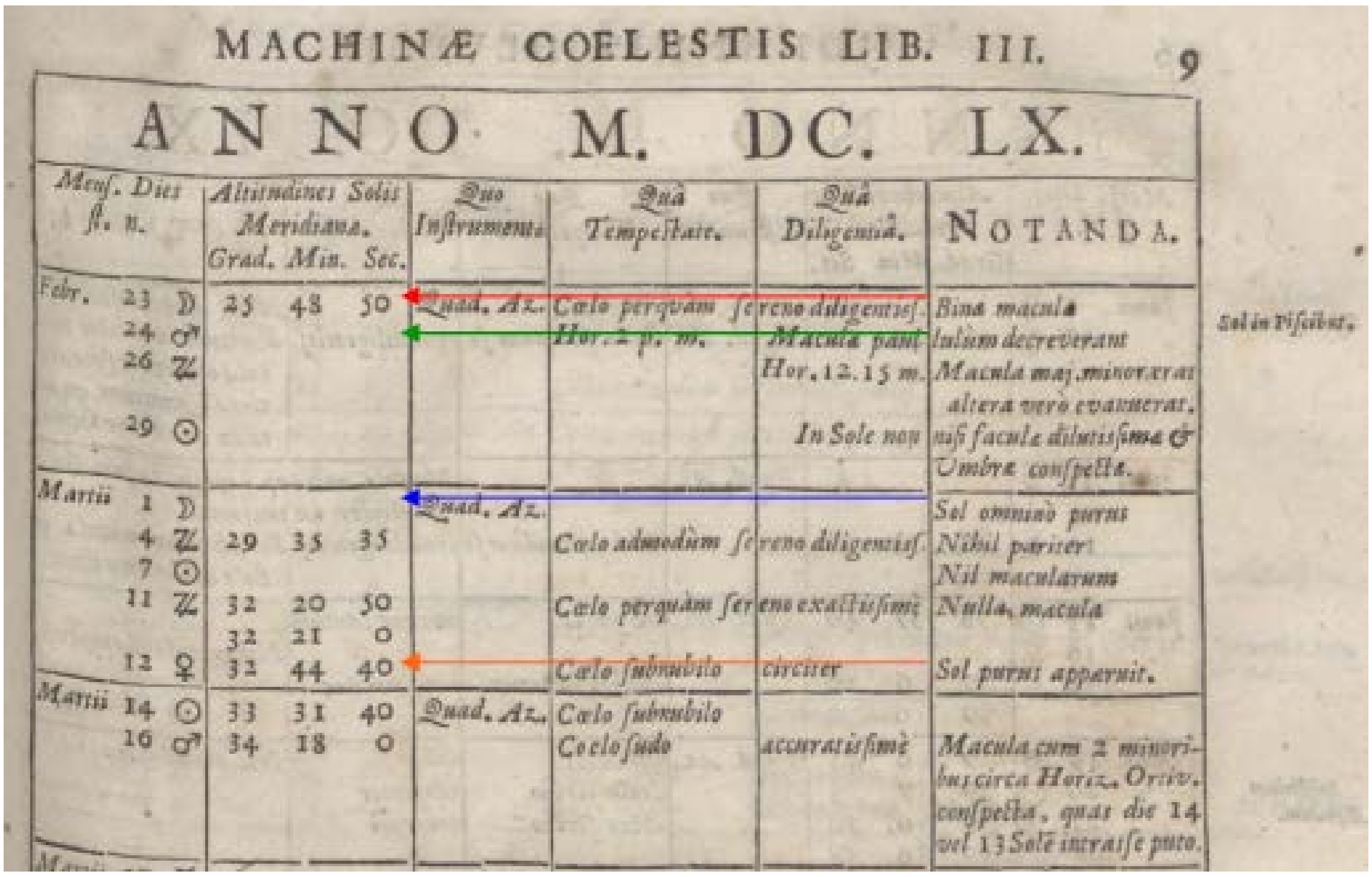

**Figure 2.** An example page of the book *Machina Coelestis* with different cases of the association between solar disc observations and solar meridian altitude measurements.

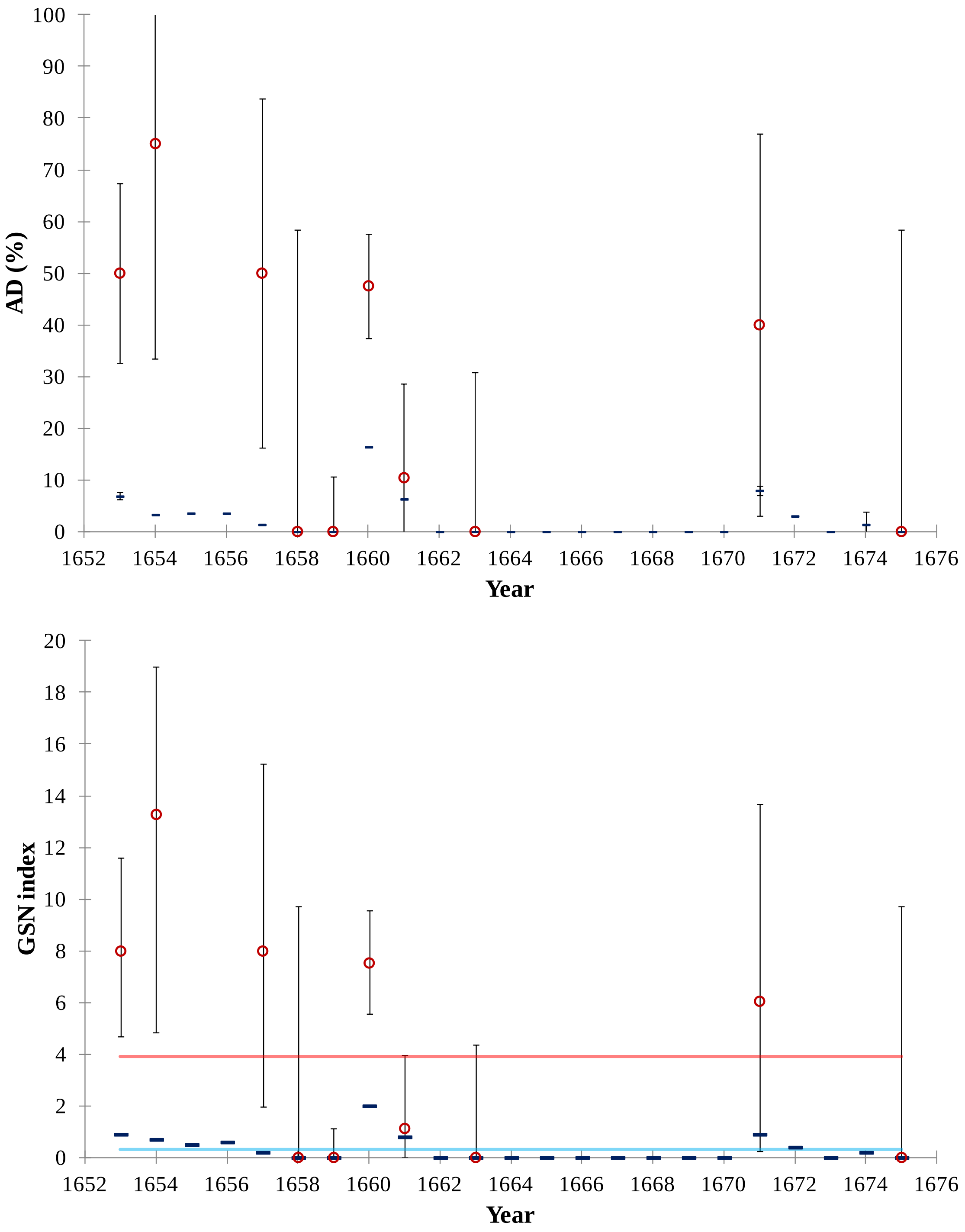

**Figure 3.** Temporal evolution of active days (%) (upper panel) and GSN (lower panel) during the period 1653-1675. Red (blue) symbols represent the values obtained in this work (by Hoyt and Schatten, 1998). The red (blue) solid line represents the mean value of GSN obtained in this work (by Hoyt and Schatten, 1998). Error bars represent a 99% confidence interval.

**Appendix. Sunspot Observations from *Machina Coelestis* by Hevelius.**

| Year | Date | Original comment [Translation] | Active Day (AD) or Quiet Day (QD) |
|---|---|---|---|
| **1653** | Feb. 5 | *Macula exigua apparuit* [Small sunspot appeared] | AD |
| | Mar. 1 | *Nil Macularum* [No spot] | QD |
| | Mar. 9 | *Nil Macularum* [No spot] | QD |
| | Mar. 23 | *Macula in quadrante Orientali Solis* [Sunspot in the eastern quadrant of the Sun] | AD |
| | Mar. 24 | *Macula Solis* [Sunspot] | AD |
| | Mar. 25 | *Binae maculae in Sole* [Double spot in the Sun] | AD |
| | Mar. 27 | *Iam penitus disparuerant* [They had completely disappeared] | QD |
| | Mar. 29 | *Nil Macularum* [No spot] | QD |
| | Apr. 6 | *Nil Macularum* [No spot] | QD |
| | Apr. 8 | *Nihil prorsus apparuit* [No spot at all] | QD |
| | Apr. 11 | *Nihil* [Nothing] | QD |
| | Apr. 13 | *Nihil* [Nothing] | QD |
| | May 20 | *Hucusque nihil Macularum in Sole apparuit* [So far no spots appeared on the Sun] | QD |
| | Jun. 27 | *Huc usque nihil in Sole apparuit, nisi unica exigua macula* [So far nothing has appeared on the Sun, except | QD |
| | | a small spot] | |
| | Jul. 13 | *Macula rotunda circa ortum animadversa* [a round spot sighted near the ortho] | AD |
| | Jul. 14 | *Macula adhuc major* [A spot even greater] | AD |
| | Jul. 16 | *Macula eadem cum alia debiliori, et comitatu quarundam debilissimarum* [The same spot with another weaker, accompanied by other very weak] | AD |
| | Jul. 17 | *Macula adhuc in Sole* [Still spot in the Sun] | AD |
| | Jul. 18 | *Macula densior cum comitatu suo apparuit, sed inferior disparuerat* [It has appeared a denser spot with its accompaniment, but the spot below disappeared] | AD |
| | Jul. 27 | *Nil Macularum* [No spot] | QD |
| | Aug. 6 | *Tres Maculae minutissimae in sole circa Ortum* [Three very tiny spots on the Sun, next to the ortho] | AD |
| | Aug. 8 | *Septem Maculae in una congerie conspectae. Septima vero multum in Sole erat promotior* [Seven spots sighted in a single mass. The seventh had increased significantly on the sun] | AD |
| **1654** | Aug. 25 | *Duae Maculae in Sole* [Two spots on the Sun] | AD |
| | Aug. 27 | *Maculae iam exiverant* [The spots were gone] | QD |
| | Sep. 18 | *Tres insignes maculae circa horizontem Ortivum conspectae* [Three obvious spots observed near the eastern limb] | AD |
| | Sep. 19 | *Quatuor Maculae circa Solis centrum deprehensae* [Four spots discovered near the center of the Sun] | AD |
| **1657** | Dec. 17 | *Nil Macularum* [No spot] | QD |
| | Dec. 21 | *Nil Macularum* [No spot] | QD |
| | Dec. 22 | *Macula exigua rotunda* [Small round spot] | AD |
| | Dec. 23 | *Macula decrescens* [The spot is decreasing] | AD |
| | Dec. 24 | *Magis magisque decrescit* [Increasingly decreases] | AD |
| | Dec. 26 | *Macula, quae in sole disco orta est, occidit Physice* [The spot that had appeared on the disc of the Sun has disappeared physically] | QD |
| **1658** | Nov. 3 | *Nulla macula* [No spot] | QD |
| | Nov. 9 | *Nil macularum* [No spot] | QD |
| **1659** | Feb. 23 | *Nulla macula* [No spot] | QD |
| | Feb. 24 | *Nil penitus macularum* [No spot at all] | QD |
| | Feb. 26 | *Nil macularum* [No spot] | QD |
| | Feb. 27 | *Nil macularum* [No spot] | QD |
| | Mar. 5 | *Nulla macula* [No spot] | QD |
| | Mar. 16 | *Nihil* [Nothing] | QD |
| | Mar. 17 | *Nihil penitus* [Completely nothing] | QD |
| | Mar. 23 | *Nil macularum* [No spot] | QD |
| | Mar. 29 | *Nil prorsus* [Nothing at all] | QD |
| | Mar. 30 | *Nulla macula* [No spot] | QD |
| | Apr. 6 | *Nullae maculae* [No spots] | QD |
| | Apr. 10 | *Nil macularum* [No spot] | QD |
| | Apr. 12 | *Nihil* [Nothing] | QD |
| | Apr. 16 | *Nihil* [Nothing] | QD |
| | Apr. 26 | *Nulla macula* [No spot] | QD |
| | Apr. 30 | *Nil macularum* [No spot] | QD |
| | May 4 | *Sol purus omnino* [The Sun completely is pure] | QD |
| | May 7 | *Nil macularum* [No spot] | QD |
| | May 11 | *Nil macularum* [No spot] | QD |
| | May 13 | *Nil macularum* [No spot] | QD |
| | May 14 | *Nil penitus* [Completely nothing] | QD |
| | May 18 | *Nil macularum* [No spot] | QD |
| | May 24 | *Nihil penitus in Sole* [Completely nothing on the Sun] | QD |
| | May 25 | *Nihil in Sole apparuit* [Nothing appeared on the Sun] | QD |
| | May 27 | *Nil macularum* [No spot] | QD |
| | May 29 | *Nil prorsus deprehensum* [Absolutely nothing is observed] | QD |
| | Jun. 2 | *Nil macularum* [No spot] | QD |
| | Jun. 4 | *Nihil* [Nothing] | QD |
| | Jun. 6 | *Nulla macula* [No spot] | QD |
| | Jun. 13 | *Sol mundus* [The Sun is clear] | QD |
| | Jun. 16 | *Nil macularum* [No spot] | QD |
| | Jun. 19 | *Nil in Sole* [Nothing on the Sun] | QD |
| | Jun. 20 | *Sol purus* [The Sun is pure] | QD |
| | Jun. 22 | *Nulla macula* [No spot] | QD |
| | Jun. 30 | *Nihil in Sole animadverti* [I have not noticed anything on the Sun] | QD |
| | Jul. 9 | *Nil in Sole extitit* [Nothing appeared on the Sun] | QD |
| | Jul. 12 | *Sol omnino purus* [The Sun completely is pure] | QD |
| | Jul. 20 | *Nil macularum* [No spot] | QD |
| | Aug. 9 | *Sol vacuus* [The Sun is empty] | QD |
| | Aug. 11 | *Nulla macula* [No spot] | QD |
| | Aug. 17 | *Nulla macula penitus* [No spot at all] | QD |
| | Aug. 27 | *Nihil deprehensum* [Nothing is observed] | QD |
| | Aug. 29 | *Omnino nihil apparuit* [Nothing at all appeared] | QD |
| | Sep. 14 | *Nil macularum* [No spot] | QD |
| | Sep. 22 | *Nulla macula in Sole* [No spot on the Sun] | QD |
| | Nov. 7 | *Nil macularum* [No spot] | QD |
| | Nov. 8 | *Nil macularum* [No spot] | QD |
| **1660** | Feb. 22 | *Notabilis macula rotunda cum minori in medio Solis disco apparuit, qualem vix memini ab aliquot annis me observasse* [Notable round spot appeared in the middle of the solar disc with a smaller one, like another that I remember observing a few years ago] | AD |
| | Feb. 23 | *Binae maculae* [Double sunspot] | AD |
| | Feb. 24 | *Hora 2 p.m. Maculae paullulum decreverant* [2 pm. Spots slightly resolved] | AD |
| | Feb. 26 | *Hor. 12. 15 m. Macula maior minor erat altera vero evanuerat* [12 hours 15 minutes. The greater spot had reduced. However, the other spot had disappeared] | AD |
| | Feb. 29 | *In Sole non nisi faculae dilutissimae et umbrae conspectae* [On the Sun just very faint faculae and umbrae are observed] | AD |
| | Mar. 1 | *Sol omnino purus* [Sun completely pure] | QD |
| | Mar. 4 | *Nihil pariter* [Nothing similar] | QD |
| | Mar. 7 | *Nil macularum* [No spot] | QD |
| | Mar. 11 | *Nulla macula* [No spot] | QD |
| | Mar. 12 | *Sol purus apparuit* [Sun appeared pure] | QD |
| | Mar. 16 | *Macula cum 2 minoribus circa Horizontem Ortivum conspecta, quas die 14 vel 13 Solem intrasse puto* [A spot together with two smaller observed near the eastern limb. I think both entered the Sun on the 14th or 13th] | AD |
| | Mar. 17 | *Macula hesterna maior creverat, quam aliae quinque minores dilutissimae sequebantur* [The major spot of yesterday had grown. It was followed by five other smaller and very faint] | AD |
| | Mar. 19 | *Maculae 5 in Sole* [Five spots on the Sun] | AD |
| | Mar. 20 | *Maculae 4* [Four spots] | AD |
| | Mar. 21 | *Maculae 4, in Quadrante Occidentali Solis* [Four spots in the western quadrant of the Sun] | AD |
| | Mar. 22 | *Maculae dilutiores quatuor* [Four spots fainter] | QD |
| | Mar. 28 | *Nil penitus macularum* [No spots at all] | QD |
| | 3 Apr. | *Nulla macula* [No spot] | QD |
| | 11 Apr. | *Nil in Sole apparuit* [Nothing appeared on the Sun] | QD |
| | 20 Apr. | *Nihil in Sole conspectum* [Nothing is observed on the Sun] | QD |
| | 26 Apr. | *Nil macularum* [No spot] | QD |
| | 27 Apr. | *Nulla penitus macula* [Utterly without spots] | QD |
| | 28 Apr. | *Nihil prorsus* [Absolutely nothing] | QD |
| | 12 May | *Insignis macula rotunda circa centrum Solis observata quam Solem intrasse puto die 6 circa* [Conspicuous round spot near the center of the Sun. I think it entered the Sun about day 6] | AD |
| | 13 May | *Macula centrum Solis iam transierat* [The spot has already passed the center of the Sun] | AD |
| | 14 May | *Macula fere eadem, sed ulterius promota more solito videbatur* [Spot is almost the same, but further advanced than usual] | AD |
| | 15 May | *Macula minor ac Horizonti Occidentali vicinior* [Spot smaller and closer to western horizon] | AD |
| | 16 May | *Macula haud procul Horizonte Occidentali* [Spot not far western horizon] | AD |
| | 18 May | *Macula circa Horizontem Occidentalem* [Spot close to wertern horizon] | AD |
| | 19 May | *Macula iam Solem exiverat* [Spot already gone out from the Sun] | QD |
| | 20 May | *Nil macularum* [No spot] | QD |
| | 21 May | *Sol plane purus* [Sun clearly pure] | QD |
| | 27 May | *Nihil apparuit* [Nothing appeared] | QD |
| | 2 Jun. | *Nihil deprehensum* [Nothing was discovered] | QD |
| | 10 Jun. | *Quatuor notabiles maculae observatae sunt, tres circa solis centrum, quarta in quadrante Occidentali* [Four notable spots are observed, three spots close to the | AD |
| | | center of the Sun, one spot in the western quadrant] | |
| | 11 Jun. | *Maculae numero 5 conspectae sunt, sed 2 debiliores extiterunt* [Five spots were sighted, but two were weaker] | AD |
| | 12 Jun. | *Debilissimae maculae in Sole adhuc extiterunt* [The weakest spots were still on the Sun] | AD |
| | 16 Jun. | *Nil macularum apparuit* [No spots appeared] | QD |
| | 19 Jun. | *Nil macularum* [No spot] | QD |
| | 9 Jul. | *Macula insignis rotunda cum alia minore Australem Circa Horizontem Occidentalem apparuit, quam antes dies circiter 10 Solis discum ingressam esse puto* [Great round spot appeared near the south-western horizon with another smaller spot that entered the solar disc about 10 days before, I think] | AD |
| | 12 Jul. | *Maculae istae iam plane evanuerant. Sed alia longe minor, cum adhaerentibus aliis quibusdam minutisimis umbrisque conspecta circa Horizontem Ortivum, quae pridie Solis discum intraverat* [The spots have clearly disappeared. But another smaller with a few others adhered to it and very tiny with umbrae is observed near the ortho horizon, which had entered the disc of the Sun the day before] | AD |
| | 13 Jul. | *Maculae sere adhuc debiliores* [The spots become weaker in the afternoon] | AD |
| | 19 Jul. | *Nulla macula* [No spot] | QD |
| | 22 Jul. | *Nil macularum* [No spot] | QD |
| | 27 Jul. | *Nihil prorsus in Sole* [Absolutely nothing on the Sun] | QD |
| | 28 Jul. | *Macula rotunda prope Horizontem Ortivum deprehensa, quae die hesterna primum Solem ingressa est* [Round spot detected near the ortive horizon, which on the day before entered on the Sun for the first time] | AD |
| | 29 Jul. | *Macula paullo creverat circa Horizontem Ortivum* [The spot had grown a bit near the ortho horizon] | AD |
| | 31 Jul. | *Macula non procul centro Solis commorabatur* [Spot stayed not far from the center of the Sun] | AD |
| | 3 Aug. | *Macula in Quadrante Occidentali* [Spot in occidental quadrant] | AD |
| | 6 Aug. | *Macula prope Horizontem Occidentalem longe minor apparuit* [A much smaller spot has appeared near the western horizon] | AD |
| | 7 Aug. | *Macula in Sole adhuc conspecta* [Spot still on the sun] | AD |
| | 23 Aug. | *Nulla macula* [No spot] | QD |
| | 28 Aug. | *Nil macularum* [No spot] | QD |
| | 24 Oct. | *Nulla macula* [No spot] | QD |
| | 2 Nov. | *Nil macularum* [No spot] | QD |
| | 11 Nov. | *Hora pom. 1 et 2 nihil quicquam in Sole visum* [1 pm and 2 pm, nothing whatever has been seen on the Sun] | QD |
| | 26 Nov. | *Nulla macula* [No spot] | QD |
| | 27 Nov. | *Nil macularum* [No spot] | QD |
| | 29 Nov. | *Nil macularum* [No spot] | QD |
| **1661** | Feb. 3 | *Macula in Sole circa centrum Solare* [Spot in the Sun, close to its center] | AD |
| | Feb. 5 | *Macula Solis* [Sunspot] | AD |
| | Feb. 10 | *Macula iam exiverat* [Spot is gone] | QD |
| | Mar. 2 | *Nil macularum* [No spot] | QD |
| | Mar. 10 | *Nihil prorsus* [Nothing at all] | QD |
| | Mar. 11 | *Nil macularum* [No spot] | QD |
| | Mar. 14 | *Nulla macula* [No spot] | QD |
| | Mar. 16 | *Nil macularum* [No spot] | QD |
| | Mar. 20 | *Nil macularum* [No spot] | QD |
| | Mar. 21 | *Nulla macula* [No spot] | QD |
| | Apr. 16 | *Nil macularum* [No spot] | QD |
| | Apr. 21 | *Nulla macula* [No spot] | QD |
| | May 1 | *Nil prorsus macularum* [No spot at all] | QD |
| | May 2 | *Nil macularum* [No spot] | QD |
| | May 3 | *Nulla quidem macula, sed Mercurius in disco Solis conspectus* [No spot, but Mercury is observed on the solar disc] | QD |
| | May 4 | *Nil quicquam macularum* [No spot] | QD |
| | May 23 | *Nil macularum* [No spot] | QD |
| | Jun. 3 | *Nulla macula* [No spot] | QD |
| | Jul. 22 | *Nil macularum* [No spot] | QD |
| **1663** | Mar. 15 | *Nil Macularum* [No spot] | QD |
| | Mar. 16 | *Nil Macularum* [No spot] | QD |
| | Mar. 17 | *Nil penitus in Sole* [Completely nothing on the Sun] | QD |
| | Mar. 19 | *Nil macularum* [No spot] | QD |
| | Apr. 2 | *Nil macularum* [No spot] | QD |
| | Apr. 3 | *Nil macularum* [No spot] | QD |
| | Apr. 4 | *Nihil prorsus* [Nothing at all] | QD |
| **1671** | Aug. 9 | *Macula Solis notabilis conspecta Hamburgi a Cl. Picardo die 13. Stetini vero a die 8 ad 18, quam etiam Lipsiae viderunt* [A clear sunspot observed by Cl. Picardo in Hamburg on the 13th. Stetini watched it from day 8 to day 18, which also was seen in Leipzig] | AD |
| | Sep. 5 | *Macula in Sole* [Sunspot on the Sun] | AD |
| | Sep. 16 | *Nulla Macula in Sole* [No sunspot on the Sun] | QD |
| | Sep. 30 | *Nulla in Sole Macula* [No sunspot on the Sun] | QD |
| | Oct. 2 | *Nihil in Sole apparuit* [Nothing appeared on the Sun] | QD |
| **1675** | Jun. 22 | *Nulla Macula* [No spot] | QD |
| | Jun. 23 | *Nulla Macula* [No spot] | QD |